\documentclass[trackchanges,twocolumn]{aastex701}
\usepackage{multirow}
\usepackage{anyfontsize}

\begin{document}

\title{Predictive Rankings of the Probability for Temperate Terrestrial Worlds for the HWO ExEP Mission Star List}

\author[0000-0001-6320-7410]{Jamie Dietrich}
\affiliation{School of Earth and Space Exploration, Arizona State University, 781 Terrace Mall, Tempe, AZ 85287}
\email{jdietrich@asu.edu}

\author[0009-0008-3554-7205]{Katelyn Ruppert}
\affiliation{School of Earth and Space Exploration, Arizona State University, 781 Terrace Mall, Tempe, AZ 85287}
\email{krupper1@asu.edu}

\author[0000-0003-2828-0334]{Austin Ware}
\affiliation{School of Earth and Space Exploration, Arizona State University, 781 Terrace Mall, Tempe, AZ 85287}
\email{atware@asu.edu}

\author[0000-0003-1705-5991]{Patrick A Young}
\affiliation{School of Earth and Space Exploration, Arizona State University, 781 Terrace Mall, Tempe, AZ 85287}
\email{patrick.young.1@asu.edu}

\begin{abstract}

The Habitable Worlds Observatory (HWO) is NASA's flagship mission design from the Decadal Survey on Astronomy and Astrophysics 2020, meant to observe temperate terrestrial planets via direct imaging and use direct spectroscopy of exoplanet reflected light to investigate their atmospheres for biosignatures. However, there are no known stars in the solar neighborhood conducive to direct imaging observations that are currently known to host rocky planets in their circumstellar habitable zones. Thus, HWO will most likely be running a blind survey; however, prioritizing the rankings of its target stars will help to potentially increase the yield of temperate terrestrial planets observed. Here we use simulated planetary systems with both small and giant planets to test which stellar systems among the HWO Exoplanet Exploration Program (ExEP) Mission Star List are most likely to host a rocky planet with the right temperature to sustain life on its surface. Assuming a simple model of planetary systems with small planets well-ordered in period interior to giant planets based on their respective occurrence rates, we find that some systems are upwards of 50\% likely to host a temperate terrestrial planet. We also consider the possibility of a giant planet in or just beyond the circumstellar habitable zone that could host a temperate terrestrial moon capable of hosting life. Additional observations to refine the occurrence rates of small planets at orbital distances $\gtrsim$1 AU and conditional rates between small and giant planets will refine these analyses and provide updates to these rankings.

\end{abstract}



\section{Introduction}{\label{sec:intro}}

The purpose of the recommended Habitable Worlds Observatory (HWO), per the 2020s Astrophysics Decadal Survey ``Pathways to Discovery in Astronomy and Astrophysics", is to ``identify and characterize Earth-like planets outside this solar system, with the ultimate goal of obtaining imaging of potentially habitable worlds" \citep[][]{Astro2020}. To facilitate precursor science for HWO, the NASA Exoplanet Exploration Program curated a list of possible target stars, the HWO ExEP Mission Star List (EMSL), that consists of 164 of the nearest single FGK stars. These targets were then ranked via initial tiers based on potential mission design requirements, including e.g., brightness, angular resolution of the HZ compared to the possible inner working angle (IWA) of the spacecraft, presence of a debris disk, etc. \citep[][]{Mamajek2024}. An additional, more comprehensive catalog was subsequently created, consisting of nearly 13,000 nearby bright stars called the HWO Preliminary Input Catalog \citep[HPIC;][]{Tuchow2024}. The HWO Living Worlds Working Group and its Target Stars and Systems subgroup created a new ranking system, reclassifying all the target stars from the EMSL as Tier 1, the entire HPIC catalog as Tier 3, and an intermediate group of targets within HPIC but not the EMSL as Tier 2 \citep[][]{Tuchow2025}.

To date, we have discovered and confirmed over 6,000 exoplanets\footnote{NASA Exoplanet Archive, 14 April 2026}, utilizing dozens of different surveys and observational techniques. The most robust population of planets we have measured comes from the Kepler Space Telescope \citep[][]{Borucki2010}, which in its primary mission discovered thousands of small ($R \lesssim 4-6~R_{\oplus}$) planets with orbital periods less than 2 years around FGK stars \citep[e.g.,][]{Burke2015,Mulders2018,Weiss2018}. Having a well-measured population of planets from one survey (with its biases and sensitivity well-determined) is useful, but we would also like to combine the demographics from different surveys to understand the planet population on a much larger scale. Current analyses are using Kepler, the radial velocity California Legacy Survey \citep[CLS;][]{Fulton2021}, and the direct imaging SPHERE infrared survey for exoplanets \citep[SHINE;][]{Desidera2021} to study giant planet occurrence rates across four orders of magnitude in orbital separation \citep[e.g.][G. J. Bergsten et al., submitted.]{Clanton2014,Clanton2016}. Conditional occurrence rates of specific types of planets have also been studied, including specifically the presence of outer giant companions to smaller inner planets and how outer giants might affect the extent of inner systems for detectability with HWO \citep[e.g.,][]{Zhu2018a, Bryan2019, Rosenthal2022, Millholland2022, Sagynbaeva2025}. However, true occurrence rates measured within exoplanet systems remain not fully constrained.

Explorations of any potentially habitable worlds pre-HWO, though, are very unlikely. The ELTs projected to come online in the 2030s might be able to find and then characterize a planet in the habitable zone (HZ) around a few of the nearest stars \citep[][]{Hardegree-Ullman2023}, but HWO will likely be running a relatively blind survey. Therefore, prioritized rankings of the systems most likely to host an Earth-like planet based on current occurrence rates would improve the detection efficiency of HWO. \citet[]{Kane2024} studied this from a purely dynamical perspective, inserting an Earth-sized planet at various places within the HZ of the 30 stars on a preliminary target list with known planets at the time, and produced a calculation of the probability the system remains stable with the additional planet. Their study provides a strong starting point for this kind of precursor analysis, but doesn't cover the probability of actually finding such a planet in the system at those locations. Additional studies by \citet[][]{Harada2024,Harada2025} studied the stellar properties and archival radial velocity data for the EMSL stars. Their analysis provided stellar companion minimum mass limits, revised and updated planet properties for known planets, and measures of the stellar activity in the host stars, an important property that was not explicitly factored into the construction of the EMSL. In particular, \citet[][]{Harada2025} note that the typical star in the list did not exhibit strong stellar activity, but over two dozen separate RV signals of stellar activity, including S-index, $R'_{HK}$, and stellar rotation periods, were found across different stars in the EMSL.

For systems with known planets, we can make testable hypotheses for the presence of potentially habitable planets using \texttt{DYNAMITE} \citep[][]{Dietrich2020}. \texttt{DYNAMITE} is a software package created to combine the specific yet incomplete data we have on a given exoplanet system with statistical models derived from the most robust population of exoplanets currently known to predict the presence and parameters of an additional planet. It uses a dynamical stability criterion \citep[][]{Dietrich2022, Dietrich2024a} and incorporates observational non-detection limits \citep[][]{Dietrich2024b} to further increase the accuracy of the predictions. The models underpinning the analysis done by \texttt{DYNAMITE} are derived from the \textit{Kepler} population of small planets, with orbital period ranges of $0.5-730$ days and $0.5-6\,R_\oplus$.

It was tested on the suite of multi-planet systems found by TESS by 2020 \citep[][]{Dietrich2020}, with a low success rate for the original 2020 predictions \citep[][]{Turtelboom2025}. However, those predictions have since been enhanced by new models and observational constraints to arrive at $\sim$50-70\% accuracy on the parameters of the 31 additional planets that were found within the previous five years in those systems \citep[][]{Dietrich2023}. We note that the prevalence of Kepler-like systems around FGK stars ranges anywhere from 30-80\% \citep[e.g.,][]{Mulders2018, Zhu2018b, Zhu2020, Yang2020}, yet this is still the most robust population of planets known, and one that also does not preclude the possibility of a temperate terrestrial planet. Therefore, we believe this is a solid base population for which to test target rankings. \texttt{DYNAMITE} was also tested on three interesting nearby systems with the potential to host an Earth-like planet that are part of the HWO EMSL: $\tau$ Ceti \citep[][]{Dietrich2021}, HD 219134 \citep[][]{Dietrich2022} and HD 20794 \citep[][]{Basant2022}. In both the original analyses and in the re-evaluation done by \citet[][]{Dietrich2024b} including observational upper limits on planet non-detections, the three systems all had a significant non-zero probability ($\gtrsim50\%$) that an additional detected planet would be a terrestrial planet in the temperate zone of the star.

This paper is laid out as follows. In Section~\ref{sec:methods} we describe the methods used to quantitatively assess the 164 stellar systems from the EMSL for their probability of hosting a potentially habitable planet. We show the results of these assessments via a weighted combination of three different test metrics based on expected contributions to the potentially habitable planet probability in Section~\ref{sec:results}. In Section~\ref{sec:discussion} we discuss the implications of these types of planetary systems for ranking HWO targets, and summarize our conclusions in Section~\ref{sec:conclusions}.

\section{Methods}{\label{sec:methods}}

\subsection{Stars with no known planets}\label{subsec:simsystems}

A large majority of the systems in the EMSL (134 out of 164) are not currently known to host planets detectable by current discovery techniques\footnote{NASA Exoplanet Archive}. For these cases, we created simulated ``true" exoplanet system architectures using the following procedure (see Table~\ref{tab:occ_rates} for a full overview of the occurrence rates used):

\begin{itemize}
    \item We created a number of giant planets in the system (including periods and masses) consistent with giant planet ($M_p \gtrsim 0.1\,M_{Jup}$) occurrence rates across from 0.01-100 AU. These occurrence rates are taken from FGK stars in the Kepler population as well as RV and direct imaging surveys, including detection efficiencies and sensitivity corrections to account for different biases to the population \citep[e.g.,][G. J. Bergsten et al., submitted]{Fernandes2019,Nielsen2019,Fulton2021}. We note that there are three early-M stars in the EMSL list. While the occurrence rates of planets around M stars are shown to be different than those around FGK stars \citep[see e.g.,][]{Dressing2013, Dressing2015, Hardegree-Ullman2019, Gonglewski2025, Glusman2026}, two of these systems have known planets (and so therefore do not have full simulated systems) and the third is close to the M-K spectral type boundary. Thus, we use the FGK occurrence rates.
    \item We then used uniformly random system inclinations and drew eccentricities from a Rayleigh distribution \citep[e.g.,][]{Winn2015}. If there was more than one giant planet in the system (which occurred $\lesssim10\%$ of the time), we tested the dynamical stability of the giant planets using both an empirical stability limit that varied between 5 and 8 mutual Hill radii \citep[e.g.,][]{He2019, Dietrich2024a} as well as N-body integrations with the spectral fraction analysis from \citet[][]{Volk2020}, which tests stability over millions of orbits and project out to billions of years, the expected age for all but three of the systems in the EMSL \citep[e.g.,][]{Harada2024, Ware2025}. The spectral fraction analysis was used if the system was declared stable via the empirical stability limit, and if either method finds the giant planets unstable we continued to re-draw their parameters.
    \item We then created a number of small planets in the system following the Kepler ``peas in a pod" population statistics \citep[][]{Millholland2017, Weiss2018, Mulders2018}, out to either the stability limit with the innermost giant planet or an empirical inner system truncation limit \citep[][]{Millholland2022}. The outer limit was set in relation to the orbital period of the innermost giant planet, with a subsample of systems including a cutoff as seen by \citet[][]{Millholland2022} that would still allow for additional small planets to exist in dynamically stable configurations. We implemented a version of this empirically measured cutoff by using the numerical dynamical stability threshold for the inner small planet compared to the outer giant. We didn't specifically truncate the values based on a set orbital period, but instead allowed it to vary with the period of the giant planet. Thus, the orbital period at which the inner system of small planets ends spans a large range between simulations, which can be well interior or exterior to the habitable zone. In our simulations, this value fell within the empirical limit range from \citet[][]{Millholland2022} more than 80\% of the time. Each system was given a central planet radius value from the Kepler distribution found by \citet[][]{Mulders2018}, and then within a system the actual small planet radii were drawn from a lognormal distribution following \citet[][]{He2019}. The system inclination from the giant planets was reused, and the mutual inclinations and eccentricities of the small planets were drawn from multiplicity-dependent lognormal distributions following \citet[][]{He2020}.
    \item We tested the dynamical stability of the small planets with each other, as well as with the innermost giant planet, with the pairwise mutual Hill radii cutoff and the system-wide N-body spectral fraction analysis, and continued to re-draw the small planet parameters if they were found to be unstable with each other or with the innermost giant.
\end{itemize}

\begin{table*}[ht]
    \fontsize{6.9pt}{6.9pt}\selectfont
    \centering
    {
    \renewcommand{\arraystretch}{2}
    \begin{tabular}[c]{c|c|c|c|c}
         \textbf{Occurrence rate model} & \textbf{Parameter space} & \textbf{Distribution} & \textbf{Formula / distribution parameters} & \textbf{Ref.}\\
         \hline
         \multirow{2}{*}{Giant planet orbital separation} & \multirow{2}{*}{$0.01-100$ AU} & \multirow{2}{*}{Broken power-law} & \multirow{1}{*}{$f(a)\propto a^{0.963\pm0.250},~a \leq 2.7\pm1$ AU} & \multirow{2}{*}{[1,2,3,4]}\\
         & & & \multirow{1}{*}{$f(a) \propto a^{-0.065\pm0.477},~a > 2.7\pm1$ AU} & \\
         \hline
         Giant planet mass & $0.1-20$ $M_J$ & Uniform & $f(m_p) =$ Uniform(0.1, 20) & [1,3]\\[1.5ex]
         \hline
         Giant planet eccentricities & $0-1$ & Rayleigh & $f(e) =$ Rayleigh(0.03) & [5]\\[2ex]
         \hline
         \hline
         \multirow{2}{*}{First small planet orbital period} & \multirow{2}{*}{$0.5-730$ days} & \multirow{2}{*}{Broken power-law} & \multirow{1}{*}{$f(P) \propto (P/12)^{1.6},~P \leq 12$ days} & \multirow{2}{*}{[6]}\\
         & & & \multirow{1}{*}{$f(P) \propto (P/12)^{-0.9},~P > 12$ days} & \\
         \hline
         \multirow{2}{*}{Small planet period ratios} & \multirow{2}{*}{$0.5-730$ days} & \multirow{2}{*}{Lognormal (dimensionless spacing)} & \multirow{1}{*}{$D = 2\frac{(P2/P1)^{2/3}-1}{(P2/P1)^{2/3}+1}$} & \multirow{2}{*}{[6]}\\
         & & & \multirow{1}{*}{$f(D) =$ Lognormal(-0.39, 0.18)} & \\
         \hline
         \multirow{2}{*}{Small planet radius - system} & \multirow{2}{*}{$0.5-6\,R_\oplus$} & \multirow{2}{*}{Broken power-law} & \multirow{1}{*}{$f(R) \propto (R/3.3)^{-0.5},~R \leq 3.3\,R_\oplus$} & \multirow{2}{*}{[6]}\\
         & & & \multirow{1}{*}{$f(R) \propto (R/3.3)^{-6},~R > 3.3\,R_\oplus$} & \\
         \hline
         Small planet radius - scatter & $0.5-6\,R_\oplus$ & Lognormal & $f(R) =$ Lognormal($R_s, 0.3$) & [7]\\[2ex]
         \hline
         Small planet mutual inclinations & $0-180^\circ$ & Lognormal & $f(i) =$ Lognormal($\mu_i(n_p),~\sigma_i(n_p))$ & [8]\\[2ex]
         \hline
         Small planet eccentricities & $0-1$ & Lognormal & $f(e) =$ Lognormal($\mu_e(n_p),~\sigma_e(n_p))$ & [8]\\
    \end{tabular}
    }
    \caption{The occurrence rate models and planet parameter spaces over which they are valid for the simulated exoplanet system architectures created for this analysis. $R_s$ is the system radius, and the $\mu_{[\alpha]}$ and $\sigma_{[\alpha]}$ values are dependent on the number of planets in the system. References: [1] \citet[][]{Fernandes2019}, [2] \citet[][]{Nielsen2019}, [3] \citet[][]{Fulton2021}, [4] \citet[][]{Desidera2021}, [5] \citet[][]{Winn2015}, [6] \citet[][]{Mulders2018}, [7] \citet[][]{He2019}, [8] \citet[][]{He2020}}
    \label{tab:occ_rates}
\end{table*}

We note that cross-population demographics are difficult to obtain precisely and reliably at this moment due to a variety of survey biases and sensitivities \citep[e.g.,][]{Clanton2014, Winn2022, Espinoza-Retamal2023}. Thus, these created systems are only an estimate for the true population exoplanet system architectures, yet one that is based on current demographics for each of the known populations' subsamples \citep[e.g.,][]{Mulders2018, Fulton2021}.

\begin{figure*}[ht]
    \centering
    \includegraphics[width=0.49\linewidth]{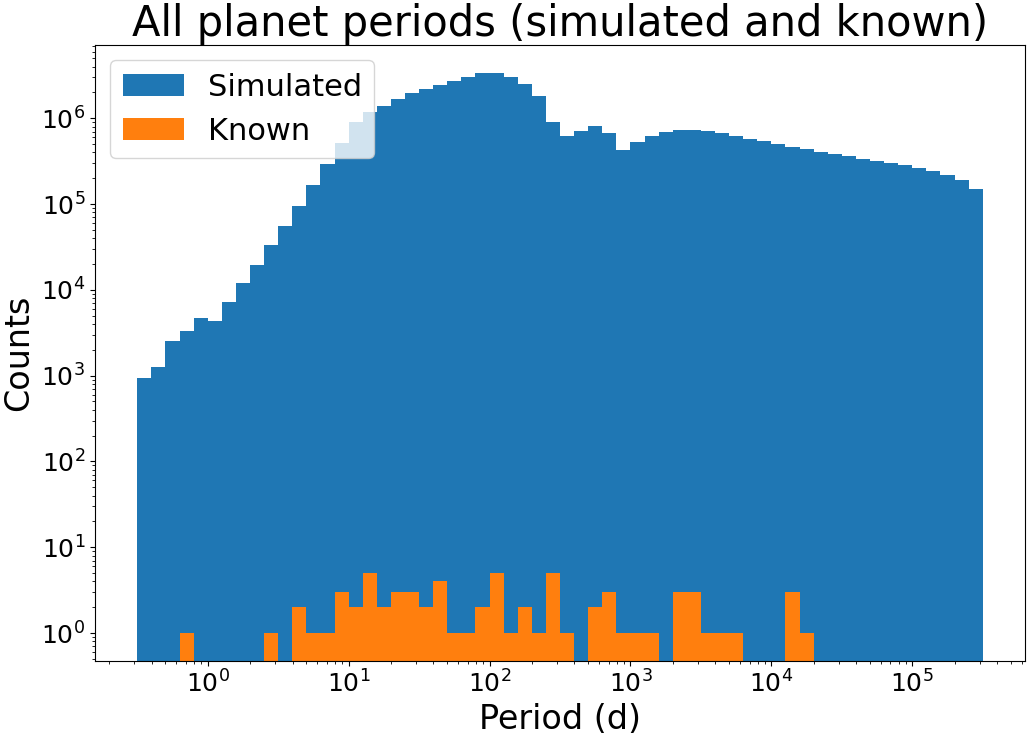}
    \includegraphics[width=0.49\linewidth]{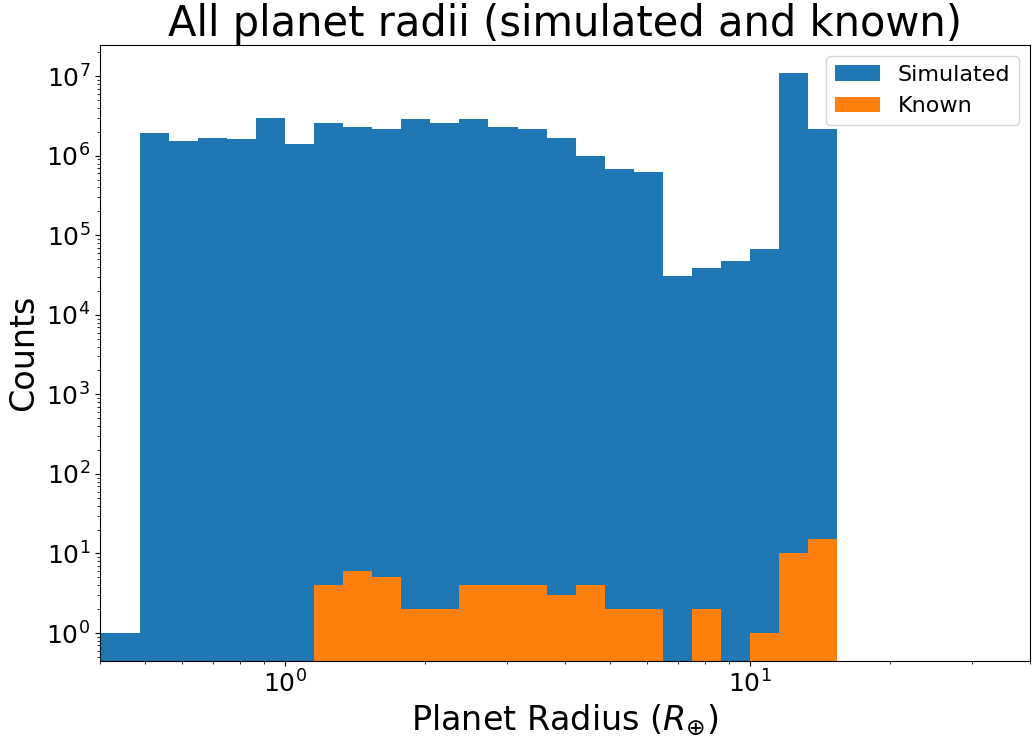}\\
    \vspace{1.5pc}
    \includegraphics[width=0.49\linewidth]{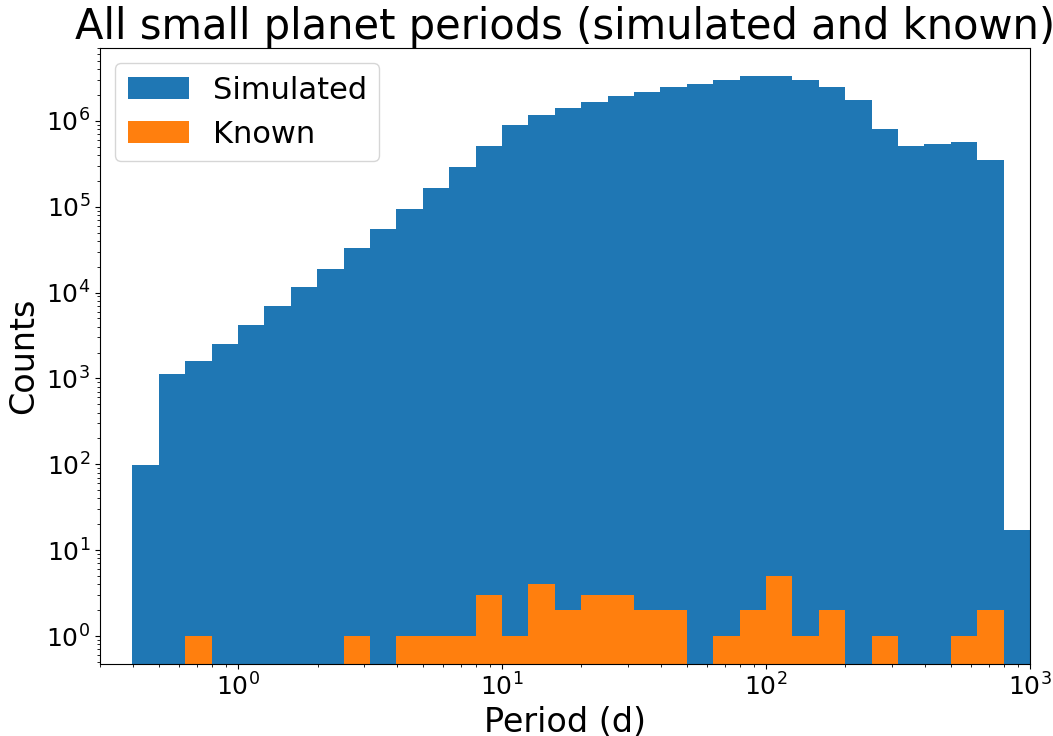}
    \includegraphics[width=0.49\linewidth]{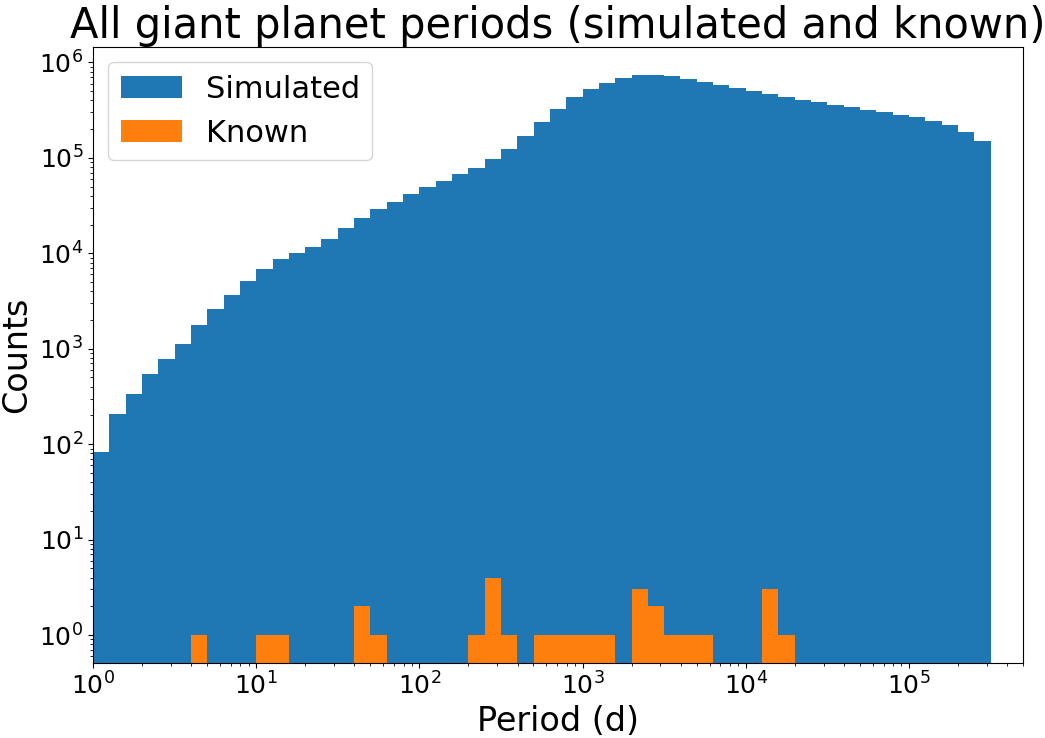}
    \caption{The simulated occurrence rates of planets in both simulated (blue) and known (orange) planetary systems around nearby stars. Top left: All planet periods. Top right: All planet radii. Bottom left: Small planet ($R<6~R_\oplus$) periods. Bottom right: Giant planet ($R>6~R_\oplus$) periods. Additional predicted planets in the systems with known planets are not shown.}
    \label{fig:dems}
\end{figure*}

Following this procedure, we used the Monte Carlo method to generate 100,000 simulated planetary systems for each star. The large range of the giant planet occurrence rates provided a large variety of simulated architectures, including hot Jupiters with no known small planets and long chains of small planets with long-period giants. The general population results are shown in Figure~\ref{fig:dems}. We use the \citet[][]{Mulders2018} occurrence rate model for the planet radius upper limit for the small planets of $6\,R_\oplus$, along with a 0.1 Jupiter mass lower limit for the giant planets to match the demographics from \citet[e.g.,][]{Fernandes2019, Fulton2021}. This mass limit is then converted into radius space using the M-R relationships from \citet[][]{Chen2017} and \citet[][]{Otegi2020}.

We see that the giant planets follow the increasing occurrence rate with period trend as expected, flattening out in the 3-5 AU range (periods between 700-3000 days), as the low sample size constraints beyond this can't exclude either upwards nor downwards trends. The small planets increase at short but increasing orbital distances \citep[periods up to 12 days;][]{Mulders2018} and then show the effects of both the truncation from inner giant planets and stability constraints based on dynamical interactions between all the planets in the system.

For each system, we then determined the fraction of simulated planets that were likely to be rocky and within the HZ. To classify planets as rocky, we considered both optimistic and conservative estimates for the radius ($R_p < \{1.5, 1.8\} ~R_\oplus$). To determine whether planets are within the HZ, we tested two sets of HZ boundaries from \cite{Kopparapu2014}: the more conservative 1 M$_\oplus$ HZ model, defined by a runaway greenhouse inner HZ and maximum greenhouse outer HZ, and the optimistic Recent Venus to Early Mars model. The authors parameterize the HZ boundaries, in terms of the instellation relative to Earth ($S_\oplus = 1361~W~m^{-2}$), with respect to the $T_{\rm eff}$ of the star. We then convert the instellations defining the inner and outer HZ to distances from the star using an inversion of the standard inverse-square law, $d = (\frac{L_\star/L_\odot}{S_\star/S_\oplus})^{1/2}$. We adopted $T_{\rm eff}$ and luminosity measurements for each star from the SPORES catalog \citep[][]{Harada2024}, which report uniformly derived stellar properties via spectral energy distribution modeling.

\subsection{Stars with known planets}

There are 30 stars in the EMSL with known planets, one of which was not analyzed by \citet[][]{Kane2024}. None of them are known to host a rocky planet in the habitable zone; the discovery methods currently available are either not well equipped to handle small planets in the temperate zone of FGK stars (direct imaging, RV for F and G stars) or require exceedingly lucky orientations of the planet and host star (transit, microlensing - also a distinctly different population). Radial velocity and astrometric observations of the EMSL targets \citep[e.g.,][]{Harada2025, Wittenmyer2025, Painter2025} have reached down to $\sim2\times$ the mass of Neptune in the habitable zone for RV and $\sim$Jupiter mass at the few-AU scale for astrometry. Also, while it is potentially feasible for TESS to find a rocky planet around a nearby star on the EMSL by gathering enough transits to raise the signal with respect to the noise, the transit probability of a temperate terrestrial planet is $<0.5\%$, thus meaning we'd need over 200 stars to expect to observe one transiting rocky planet in the habitable zone. These limitations have thus guided the search for additional planets, especially potentially habitable ones \citep[][]{Mamajek2024}. 

For the stars with known planets, we used \texttt{DYNAMITE} to predict the presence of an additional planet in the system, and we outline the functionality of \texttt{DYNAMITE} here. We start with the same set of statistical models derived using the small-planet occurrence rates as in Section~\ref{subsec:simsystems}, and the known set of parameters we utilize for each planet: the orbital period, inclination, and eccentricity, and then a parameter for the size. Planet radius is preferred since the models are derived from \textit{Kepler} data and therefore use planet radius distributions, but for mass or minimum mass $m \sin i$ we convert the values to radius for the models and then convert back when providing the estimated parameters for the new planet \citep[as in][]{Dietrich2021}. The starting set of parameters for each known planet are drawn anew for each Monte Carlo iteration of the planet system, using the provided uncertainties in their reported values as well as any observational upper limits provided \citep[][]{Dietrich2024b}). If a known planet has no inclination measurement (and therefore only has a minimum mass), we allow the inclination to be a free parameter that changes with each iteration and adjust the mass and radius each time.

\texttt{DYNAMITE} was built with multiple possible statistical models for each parameter, and we use the ones reported here due to better performance in previous tests of the system \citep[e.g.,][]{Dietrich2020, Dietrich2023, Turtelboom2025}. Here we use the period ratio distribution from \citet[][]{Mulders2018} instead of, e.g., the clustered periods distribution from \citet[][]{He2019}, but inversely the \citet[][]{He2019} clustered radius distribution instead of the \citet[][]{Mulders2018} equal intrasystem radius. The orbital mutual inclinations and eccentricities were drawn from the multiplicity-dependent Lognormal distributions from \citet[][]{He2020}. In the case of the planet radius and orbital inclinations, which are each set from a system-wide value, we fit the known small planets to get a cluster radius $R_c$ and system inclination $i_s$ such that the Lognormal distributions $f(R)$ for the planet size and $f(i_m)$ for the orbital inclinations are maximized for the values from the known planets.

For the period ratio distribution, we calculated the joint probability of the two new period ratios created for additional planets between any current pair of planets. Extrapolation beyond the outermost known planet or outside of the parameter validity range is possible, but we do not report those values for consistency. For periods interior to the first known planet, we calculate the joint probability of the new period ratio combined with the value from the first planet distribution that peaks at 12 days. Across the parameter validity range of 0.5 to 730 days, this provides the probabilities of finding a given planet with that orbital period. We then run 100,000 Monte Carlo iterations with a single planet value drawn from the posterior probability distributions we just created for all four parameters, and test the dynamical stability of the new system each time. If the additional planet were to make the system dynamically unstable, we set the probability of such a planet equal to 0. Finally, we then took the distributions of the 100,000 iterations for each parameter of the most likely planet and again determined if said iteration of the planet was likely to be rocky and within the \citet[][]{Kopparapu2014} circumstellar habitable zone boundaries. This provides a probability with uncertainties for the final results from the analysis.

\texttt{DYNAMITE} can be run iteratively to test the probability of multiple additional planets in the system \citep[see e.g.,][]{Dietrich2020, Dietrich2021, Dietrich2022, Basant2022}. We tested the possibility of finding two previously undiscovered planets in each system and if maybe the second one would be more likely to be habitable than the first; however, that was not true for any of the 30 systems with currently known planets, so we report only the probability for one additional planet in Table~\ref{tab:metrics_known}. For the systems with only one known planet, the models are less constraining than with multiple, as \texttt{DYNAMITE} tends to find a similarly-sized planet near a 2:1 orbital resonance to be the most likely, given there are little other constraints to be placed on the system. In addition, $\sim85\%$ of the known multiplanet systems extend out to or beyond the habitable zone, allowing for stricter constraints on the presence of potentially habitable planets instead of relying on extrapolated values.

Finally, the additional predicted planets share some similarities with the simulated planets, but they also have some significant differences. The predicted additional planets tend to be shorter period than the simulated planets, with the simulated planets peaking around 100 days orbital period whereas the predicted planet posteriors have a variety of peaks between 8 and 80 days. This is due both to the fact that there are 30 systems with very different known planets and each having 100,000 iterations creating their posteriors, as well as the fact that \texttt{DYNAMITE} is less constrained beyond 730 days, so the increase around 1000 days in the simulated planets is not found. The predicted additional planets tend to be larger (with a median of $2.45\,R_\oplus$ vs. $1.55\,R_\oplus$ for the simulated planets), as the known planet populations are larger than the simulated planets (mostly because of detection sensitivity biases).

\subsection{Habitable terrestrial moons of giant planets}

An additional way to increase the potential targets in the search for life is looking for terrestrial moons of giant planets that are the right temperature to host liquid water and support Earth-like life \citep[e.g.,][]{Heller2013,Hill2018,Dencs2025}. This opens up the parameter space a little further, as potential habitability based on temperature is not solely defined by the stellar insolation the body receives, but also the tidal heat flux from its orbiting planet. For every simulated giant planet created for the systems without known planets, as well as for every known giant planet in the systems currently with planets, we tested the probability for the giants hosting a habitable moon with this procedure.

\begin{itemize}
    \item We generated moons for each giant planet, with the properties set as follows: the number of moons and the average mass of the moons were modeled from simulations done by \citet[][]{Dencs2025}, with a log-uniform spacing from the Roche limit of the giant planet to 40\% of its Hill sphere \citep[a conservative limit for moon orbit stability, see e.g.,][]{Domingos2006}. The number of moons ranged from 2 to 23, and was dependent on both the mass and orbital radius of the giant planet. The average mass of the moons in a given system ranged from $0.1 M_{Luna}$ to $0.5 M_\oplus$, and was dependent on the mass of the giant planet and inversely dependent on the orbital radius of the giant planet.
    \item Each moon in the system was then given a mass drawn from a normal distribution about the modeled mean moon mass for the specific giant planet in the system, with standard deviations also modeled from \citet[][]{Dencs2025}. The moon was then given a physical radius derived from the ``rocky" portion of the M-R relationship from \citet[][]{Otegi2020}. The moons were then also given some eccentricity from a normal distribution, with its center dependent on the number of moons -- more moons meant a lower mean moon eccentricity, in order for the moon system to be dynamically stable. The initial moon eccentricities were centered around 0.15, but allowed to range up to 0.6, following the simulated average and limit for eventual accretion onto the planet from \citet[][]{Dencs2025}.
    \item We followed the procedure from \citet[][]{Dencs2025} using a simple fixed-Q model (Q being the tidal dissipation factor) from \citet[][]{Heller2013} to calculate the global heat flux of the moon including the effect of tides from the host planet.
\end{itemize}

We note that a viscoelastic model of the planetary body \citep[e.g.,][]{Henning2009, Dobos2015, Dobos2017} is preferred, but this requires an understanding of the internal structure of the body that we are extremely unlikely to be able to capture or fully model for Earth-like exomoons. For bodies like Jupiter's moon Io, the formula used by \citet[][]{Heller2013} fits its current global heat flux \citep[$2.24\pm0.45\,W/m^2$;][]{Lainey2009} better than the Maxwell model used by \citet[][]{Henning2009}, at least for internal temperatures sufficiently far from the disaggregation temperature of the body, as seen in semi-analytic estimates made using \texttt{TidalPy} \citep[][]{Renaud2023}.
    
We use the following characteristics for a moon to be considered habitable. The moon must be at least as massive as Mars so that it can retain its atmosphere, which can be lost due to solar or planetary irradiation effects for smaller masses so therefore would not be able to have a potentially habitable observable atmosphere \citep[e.g.,][]{Lammer2014, Dencs2025}. This justifies the use of the \citet[][]{Otegi2020} M-R relationships, which are valid down to Mars-mass objects, and in good agreement with values from other M-R relationships \citep[e.g.,][]{Chen2017, Muller2024}. Also, the global heat flux of a potentially habitable moon, including tides, must be within the habitable zone insolation values dependent on the mass of the potentially habitable body \citep[][]{Kopparapu2014}. This does not extend the inner edge of the circumstellar habitable zone, as the stellar flux will still make the moons too hot, but massive moons around giant planets at slightly longer distances than the circumstellar habitable zone (e.g., within 1-2 AU of a G-type star) can have the right heat flux to be potentially habitable.

\subsection{Target ranking metric}

We combined three primary metrics in order to rank the targets on the EMSL. The first criterion was the probability of the simulated and/or injected planets specifically being terrestrial planets in the habitable zone. The second was the 2 Gyr continuous HZ (CHZ$_2$) metric \citep[][]{Ware2022,Ware2025}. The CHZ$_2$ defines the orbital area around the host star that has remained within the HZ for 2 Gyr or longer, corresponding to the approximate start of the Great Oxidation Event on Earth \citep[e.g.,][]{holland2006}. \cite{Ware2025} calculate the posterior probability distribution for the location of the CHZ$_2$ and then integrate over that distribution outside a hypothetical HWO IWA to determine the CHZ$_2$ metric for each star in the EMSL. This metric essentially estimates the probability that a planet has been habitable long enough for life to have made a detectable impact on the atmosphere. The last metric was the probability of having a terrestrial moon around a giant planet within the zone where said moon would have a temperate global climate, following \citet[][]{Dencs2025}.

%The different weights were selected randomly at first, but then subjectively optimized to take into account the differences in system architecture while still allowing for time to actually create a observable biosphere. Thus we arrived at the weights of 75\% for having an Earth-like planet in the habitable zone, which we denote as $\hat{\eta_\oplus}$, 20\% for the $CHZ_2$ metric, and 5\% for having an Earth-like moon around a Jovian planet in the habitable zone.

Since the CHZ$_2$ is calculated assuming the existence of a temperate terrestrial planet, we created our ranking metric based on the sum of the first and third criteria (the probability of the existence of a temperate terrestrial planet and moon, respectively) multiplied by the second criterion. We note that the temperate terrestrial moons around giant planets do not necessarily need to exist in the current circumstellar HZ due to tidal heating, but as the HZ moves outwards with increasing stellar luminosity as the star ages, the heat flux of these moons is also affected in a similar way to temperate terrestrial planets.

\section{Results}{\label{sec:results}}

We ran both the simulated and the known systems through 100,000 Monte Carlo iterations, with the simulated systems being recreated anew each time and the known system planet parameters getting redrawn from the literature distributions without any additional planets included. For each of these iterations, we count if a simulated planet or the predicted additional planet in the known system fits the parameters for a temperate terrestrial planet. When divided by the number of iterations, this yields a probability of finding an Earth-like planet in the habitable zone for these systems. We note that since the statistical models used by \texttt{DYNAMITE} for systems with known planets do not extend out beyond the AU scale, whereas simulated planets in systems without known planets exist out to 100 AU for the integrated giant planet demographics, we consider these as two separate populations that aren't meant to be directly comparable. The results for the systems with known planets are shown in Table~\ref{tab:metrics_known}, whereas those for the systems without known planets are shown in Table~\ref{tab:metrics_unknown} in Appendix~\ref{app:simulated}.

\begin{table*}[ht]
    \centering
    \footnotesize
    \begin{tabular}{c|c|c|c|c|c|c|c}
       \textbf{System} & \textbf{Alt. Name} & $\mathbf{\hat{\eta_\oplus}}$ & $\mathbf{CHZ_2}$ & \textbf{Moon} & \textbf{\textit{Joint Metric}} & \textbf{\textit{K24 DVHZ}} & \textbf{EMSL Tier}\\
       \hline
HD 115617 & 61 Vir & \{0.4634, 0.5666\} & 0.576 & \{2e-05, 2e-05\} & 0.2966 & 44.93\% & B\\
HD 143761 & rho CrB & \{0.3102, 0.4264\} & 0.535 & \{0.0, 2e-05\} & 0.197 & 88.17\% & B\\
HD 219134 &  & \{0.4334, 0.5402\} & 0.391 & \{0.0, 0.0\} & 0.1903 & 99.26\% & A\\
HD 10700 & tau Cet & \{0.2535, 0.3299\} & 0.467 & \{0.0, 0.0\} & 0.1362 & 20.75\% & B\\
HD 75732 A & 55 Cnc A & \{0.3677, 0.5085\} & 0.256 & \{0.0, 0.0\} & 0.1122 & 56.99\% & C\\
HD 20794 & 82 Eri & \{0.1233, 0.2666\} & 0.520 & \{0.0, 1e-05\} & 0.1014 & 0\% & B\\
HD 190360 &  & \{0.1692, 0.2211\} & 0.353 & \{0.0, 0.0\} & 0.0689 & 85.67\% & B\\
HD 140901 A &  & \{0.0689, 0.098\} & 0.169 & \{0.0, 0.0\} & 0.0141 & 89.06\% & C\\
HD 39091 & pi Men & \{0.0042, 0.0084\} & 0.321 & \{0.02666, 0.04779\} & 0.014 & 0\% & B\\
HD 136352 & nu2 Lup & \{0.0183, 0.0528\} & 0.362 & \{0.0, 0.0\} & 0.0129 & 100\% & B\\
HD 160691 & mu Ara & \{0.0, 0.0\} & 0.706 & \{0.00743, 0.02214\} & 0.0104 & 5.39\% & B\\
HD 217987 &  & \{0.0677, 0.1494\} & 0.044 & \{1e-05, 2e-05\} & 0.0048 & 100\% & B\\
HD 86728 A & 20 LMi A & \{0.0007, 0.004\} & 0.554 & \{2e-05, 3e-05\} & 0.0013 & N/A & B\\
HD 209100 & eps Ind & \{0.0, 0.0\} & 0.363 & \{0.00239, 0.00392\} & 0.0011 & 100\% & A\\
HD 69830 &  & \{0.0014, 0.0076\} & 0.234 & \{3e-05, 3e-05\} & 0.0011 & 84.48\% & C\\
HD 22049 & eps Eri & \{0.0, 0.0\} & 0.370 & \{0.00204, 0.00314\} & 0.001 & 100\% & C\\
HD 3651 A & 54 Psc A & \{0.0, 0.0\} & 0.283 & \{0.00263, 0.00341\} & 0.0009 & 92.41\% & B\\
HD 102365 &  & \{0.0002, 0.0018\} & 0.567 & \{0.00013, 0.00019\} & 0.0007 & 78.21\% & A\\
HD 189567 &  & \{0.0028, 0.0094\} & 0.072 & \{1e-05, 2e-05\} & 0.0004 & 100\% & C\\
HD 95735 & GJ 411 & \{0.0393, 0.057\} & 0.007 & \{2e-05, 2e-05\} & 0.0003 & 100\% & B\\
HD 141004 & lam Ser & \{0.0001, 0.0006\} & 0.734 & \{2e-05, 3e-05\} & 0.0003 & 100\% & A\\
HD 115404 A &  & \{0.0012, 0.0016\} & 0.048 & \{0.00269, 0.00413\} & 0.0002 & 100\% & C\\
HD 95128 & 47 UMa & \{0.0, 0.0\} & 0.693 & \{0.0, 0.00041\} & 0.0001 & 27.59\% & A\\
HD 147513 &  & \{0.0, 0.0\} & 0.358 & \{1e-05, 1e-05\} & 0.0 & 0\% & A\\
HD 10647 & q1 Eri & \{0.0, 0.0\} & 0.130 & \{0.0, 4e-05\} & 0.0 & 38.65\% & C\\
HD 17051 & iot Hor & \{0.0, 0.0\} & 0.053 & \{1e-05, 1e-05\} & 0.0 & 55.02\% & B\\
HD 192310 &  & \{0.0, 0.0\} & 0.366 & \{0.0, 0.0\} & 0.0 & 19.14\% & A\\
HD 114613 &  & \{0.0, 0.0\} & 0.728 & \{0.0, 0.0\} & 0.0 & 33.43\% & C\\
HD 33564 &  & \{0.0, 0.0\} & 0.400 & \{0.0, 0.0\} & 0.0 & 61.50\% & C\\
HD 9826 A & ups And A & \{0.0, 0.0\} & 0.892 & \{0.0, 0.0\} & 0.0 & 0\% & C\\
    \end{tabular}
    \caption{The three individual metrics and the joint metric, as well as the dynamical stability ranking from \citet[][]{Kane2024} where applicable, for each of the 30 stars on the EMSL with known planets. $\hat{\eta_\oplus}$ is the probability of finding an Earth-like planet within the habitable zone in our simulations, and K24 DVHZ is the probability of an Earth twin planet's stability being dynamically viable within the habitable zone. The conservative and optimistic cases are shown separately for the probability of the temperate terrestrial planet and moon, whereas the joint metric uses the average value in its calculation. HD 86728 b \citep[][]{Gupta2025} was discovered after the analysis done by \citet[][]{Kane2024}, whereas HD 26965 b was on the \citet[][]{Kane2024} list but recently found to more likely be stellar activity \citep[][]{Burrows2024}.}
    \label{tab:metrics_known}
\end{table*}

\begin{figure}[t]
    \centering
    \includegraphics[width=\linewidth]{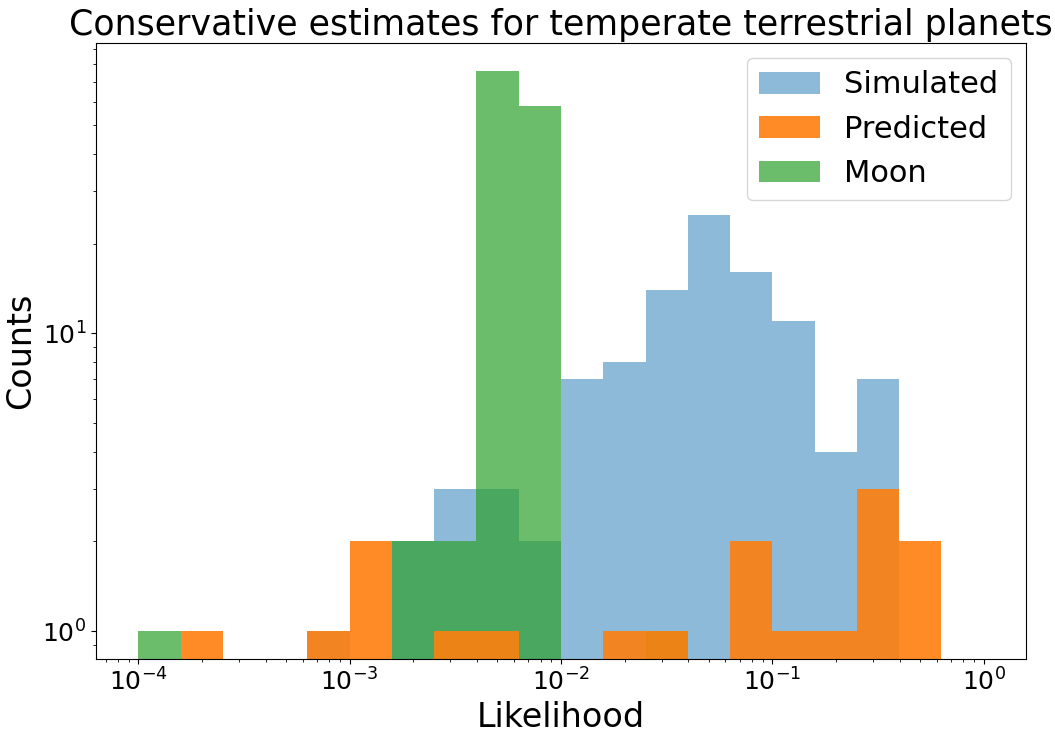}
    \includegraphics[width=\linewidth]{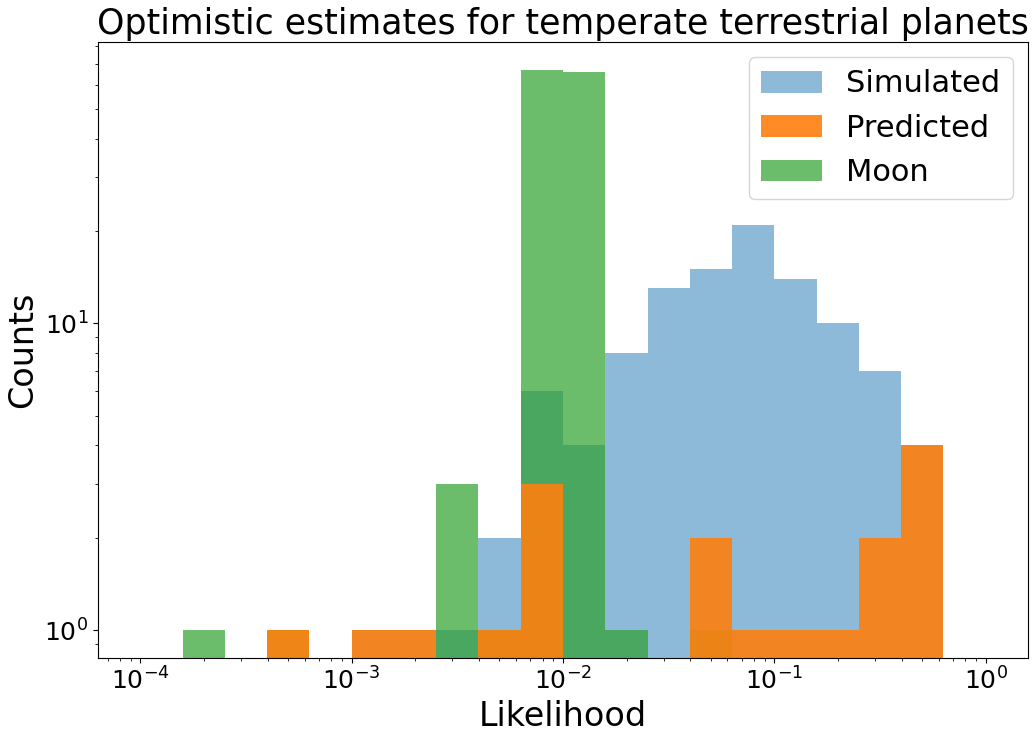}
    \caption{Probabilities of a temperate terrestrial planet for the conservative (top) and optimistic (bottom) assumptions for searching for potentially habitable worlds in nearby star systems. Simulated values are for systems with no known planets, predicted values are for systems with known planets. Systems with a probability of 0 are not shown. The number of counts in each probability bin is the number of EMSL stars that have that probability of a temperate terrestrial planet.}
    \label{fig:likes}
\end{figure}

The distribution of probabilities is shown in Figure~\ref{fig:likes} for the conservative and optimistic cases of the habitable zone. We find that there is a non-zero difference between the conservative and optimistic cases, sometimes up to a 50\% increase in relative probability for a given system. For our analysis, we use the average of the conservative and optimistic values. We also find that $\sim80\%$ of the systems with known planets and $\sim90\%$ of the systems without known planets have less than 20\% probability of hosting a rocky planet in the circumstellar habitable zone. For both sets of systems, having other small planets in a given system increases the probability of a rocky planet in the habitable zone. This matches the ``peas-in-a-pod" statistical model that Kepler planetary systems are seen to follow \citep[e.g.,][]{Millholland2017, Weiss2018}. However, since most of the known systems have larger planets at habitable zone orbital distances, and those planets have correspondingly larger mutual Hill radii, this makes the stable regions near these planets within the habitable zone significantly smaller than those in simulated systems with rocky planets. This also varies with host star to a degree, as smaller stars have closer-in yet narrower habitable zones. In general, most FGK stars can fit multiple rocky planets within the habitable zone with $\sim10$ mutual Hill radii stability \citep[like our Solar System or $\tau$ Ceti; for dynamical stability analyses of additional small planets in the habitable zone of systems like these see e.g.,][]{Dietrich2021, Dietrich2022}, yet one giant planet even outside the habitable zone but near enough to it can make most or all of the habitable zone unstable \citep[e.g.,][]{Kane2024}.

One notable difference in the results between the sample of systems with known planets vs systems with simulated planets is the frequency of temperate terrestrial planet probability equal to 0: 11 out of 30 (36.7\%) for known planets vs. 27 of 134 (20.1\%) for simulated planets. We attribute this to the fact that the known planets in a given system strongly constrain the probability distributions for those systems, since none of the currently known planets are rocky and in the circumstellar HZ. Thus, only small probabilities exist where planets are both likely to be rocky as well as dynamically stable in the circumstellar HZ with the known planets. In the scenario for systems without known planets, the overall general population of the simulated planets covers a very wide range of parameter space, allowing for non-zero probabilities of certain habitable configurations, both with temperate terrestrial planets as well as temperate terrestrial moons of giant planets.

\section{Discussion}{\label{sec:discussion}}

\subsection{Terrestrial planets in the habitable zone}

\begin{figure}[t]
    \centering
    \includegraphics[width=\linewidth]{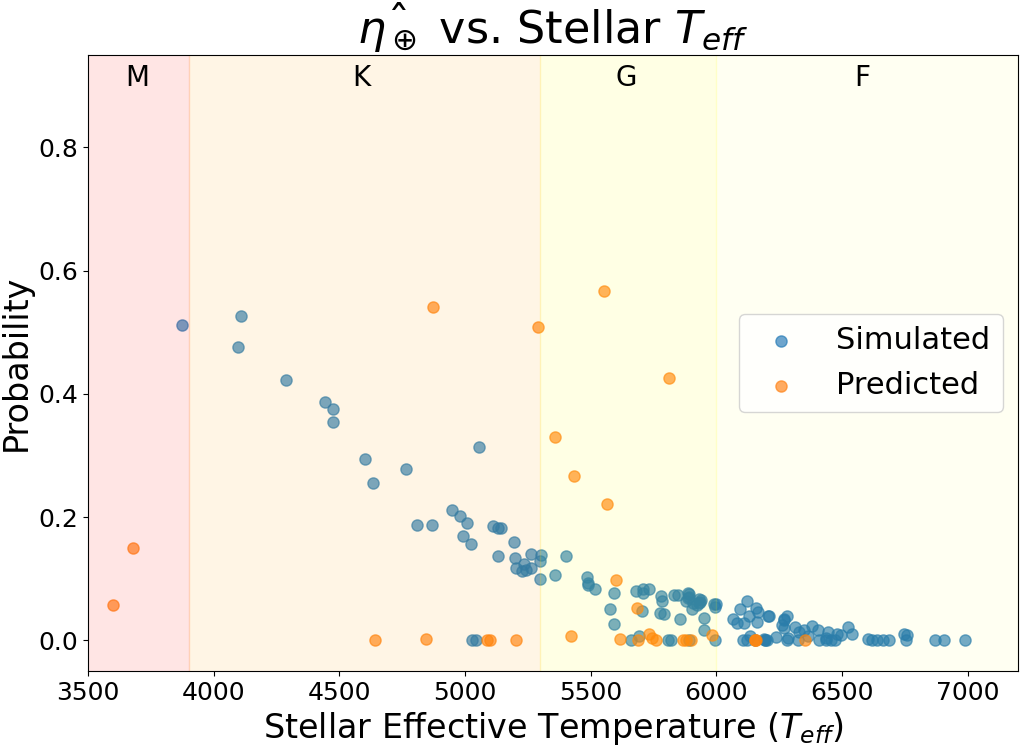}
    \caption{The simulated and predicted probability for Earth-like planets compared to the stellar effective temperature for each of the 164 targets on the HWO EMSL. The habitable zone moves outwards with increasing temperature, but the occurrence rate of giant planets also increases with orbital distance, thus limiting the simulated probability of finding small rocky planets in the habitable zones of hotter stars.}
    \label{fig:eta_steff}
\end{figure}

Here we discuss the results for the planet cases, where the known and simulated planetary systems are treated separately. The simulated systems around the stars on the HWO EMSL that are not known to contain planets are created via a simple combination of Kepler demographics for small planets and joint occurrence rates for giant planets. This tends to create smaller planets closer in to their host stars and giant planets further out, with no noticeable change in the simulated population between spectral types. However, as the habitable zones for stars moves outward with increasing temperature, the small planets are more likely to exist within the habitable zone for cooler stars but interior to the habitable zone for warmer stars. Thus, the cooler stars are more likely to contain a temperate terrestrial planet that would be a promising target for HWO, whereas the warmer stars are less likely (as shown in blue in Figure~\ref{fig:eta_steff}).

For $\sim70\%$ of the systems with known planets, the current orbital configuration of planets precludes the probability of a temperate terrestrial planet (using the Kepler demographics) to less than 1\% (known systems shown in orange in Figure~\ref{fig:eta_steff}). This does not mean there is no space for all of these systems; \citet[][]{Kane2024} found that roughly half of these systems had $>50\%$ dynamical stability for an Earth-sized planet at Earth insolation in the habitable zone. These results highlight the difference in whether a planet \textit{could} stably exist as compared to whether it is \textit{likely} to exist in a given location.

In our own Solar System, there are multiple opinions on whether and how much Jupiter affected Earth towards or away from habitability. Due to its much greater relative size (and therefore gravity), Jupiter has the largest perturbing influence on smaller bodies like terrestrial planets as well as asteroids and comets. While Jupiter does protect the Earth by taking thousands of more impacts per Earth impact and keeps the Oort Cloud comets from coming too close to Earth \citep[e.g.,][]{Lewis2013}, it also has been shown to increase the Earth impact flux for asteroids and short-period comets \citep[e.g.,][]{Horner2012, Grazier2019}. All of these combine to create some effect on Earth's ability to host life via surviving without too many impacts but also having water from ice-rich comets deposited in formation. In addition, Jupiter is theorized to be the reason why there are no planets interior to Mercury's orbit, which is in the 98th percentile of systems for closest planet \citep[e.g.,][]{Mulders2018}. Jupiter may have caused any proto-inner planets to migrate inwards into the sun and protected the further out rocky planets by carving out gaps in the gas disk interior to its orbit, before it was migrated further outwards in tandem with Saturn \citep[e.g.,][]{Batygin2015, Srivastava2025}. While we know the presence of super-Earths and cold Jupiters are correlated \citep[e.g.,][]{Zhu2018a}, thus motivating the analysis to test systems with both small and giant planets, we still cannot truly determine what effect a giant planet has on the \textit{habitability} -- not the presence -- of a small planet.

\subsection{Earth-like moons of Jovian planets in the habitable zone}

Utilizing tidal heating, there are small windows exterior to the circumstellar habitable zone where the heat flux on a small rocky body would be sufficient for liquid water on its surface \citep[][and references within]{Dencs2025}. For systems with known giant planets in or near the circumstellar habitable zone \citep[e.g., HD 9826 A;][]{Butler1999}, as well as systems where the giant planet population's expected peak in orbital distance corresponds to the habitable zone \citep[e.g.,][]{Fernandes2019,Fulton2021}, this marginally increases the probability of observing an Earth-like atmosphere around a nearby star. However, there are additional observational issues this specific setup produces.

Direct imaging with HWO will not be able to angularly resolve a potentially habitable Earth-like moon from its Jovian host planet. This is given an optimistic setup with an expected minimum inner working angle of 65 mas and a system configuration with a host star at 5 pc, Jovian planet at 2 AU orbital separation, and planet-moon separation of 0.03 AU, which provides a maximum separation of 6 mas \citep[e.g.,][]{Mamajek2024, Fernandes2025, Dencs2025}. Thus, any observation will be a spectrum of the host planet combined with the moon. A gas giant host planet will be large and bright enough to be easily observed and to get a spectrum with significantly high signal-to-noise. However, disentangling the observable signatures of a much smaller and fainter $(\lesssim1\%)$ moon compared to the planet will be possible but difficult \citep[e.g.,][]{Kleisioti2024}. The best case scenario for spectroscopic direct imaging would be to capture the planet-moon combination at a few different points in the moon's orbit around the planet, especially near quadrature, in order to catch the adjusted Doppler shift of the moon \citep[separate from the planet's own value; see e.g.,][]{Vanderburg2018}. This would require some apriori knowledge of the planet and potentially habitable moon's orbits, which is a very difficult task but nonetheless one that could be possibly be tested in the upcoming decades.

Given the above, then, an observation of a Jovian exoplanet and Earth exomoon system is likely to be similar to current gas giant model spectra, with an enhancement from $\sim0.1-1\%$ up to $\lesssim10\%$ flux ratios in the spectral lines corresponding to a terrestrial object \citep[e.g.,][]{Irwin2014}, also with a potential offset from the gas giant's lines, that would make observing a habitable exomoon possible with HWO. Notably, gas giants are likely to contain strong evidence for a few common potential biosignatures and bio-essential compounds like methane and water \citep[e.g.,][]{Macintosh2015}, so constraining the specific biosignature combinations that would be considered strong evidence for life would be necessary in order to rule out false positives due to contamination from the host planet.

\subsection{HWO target rankings}

\citet[][]{Kane2024} provided rankings on the systems with known planets by simply injecting an Earth ($R_p = R_\oplus$) into the habitable zone of each star and calculating the dynamical stability of the system and the expected long-term stability of the system with the added planet. For these systems, we find that our dynamical assessment of the most likely additional planet to add to these systems is consistent roughly half the time with an Earth-like planet as in \citet[][]{Kane2024}, with an additional $\sim30\%$ of additional planets matching a peak in the orbital period distribution but expected to be significantly larger than Earth. We also see a consensus between the EMSL Tiers and the Joint Metric for the systems without known planets, but the Tier A systems in the known planets are spread relatively equally throughout the sample while the Tier B systems are generally above the Tier C systems. 

\begin{figure}[t]
    \centering
    \includegraphics[width=\linewidth]{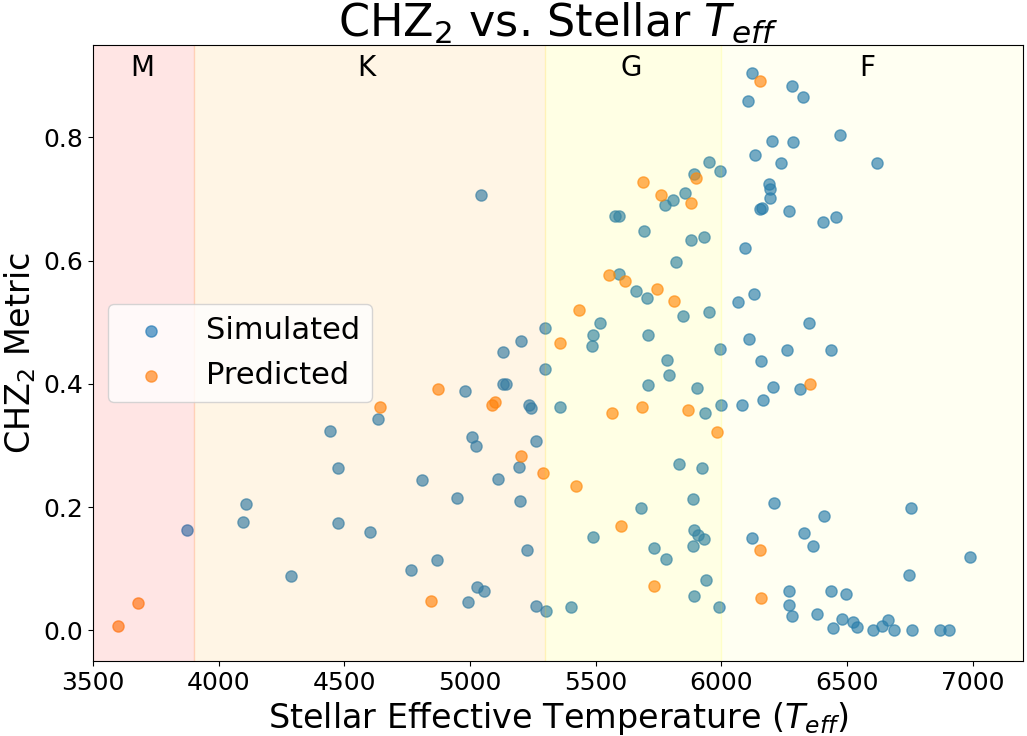}
    \caption{The CHZ$_2$ metric compared to the stellar effective temperature. In general, the CHZ$_2$ has a positive trend with stellar temperature until $\sim6500$ Kelvin, where the stars start to become too short-lived on the main sequence to have a stable circumstellar habitable zone for 2 billion years. This is in contrast to the trend of the probability in finding a temperate terrestrial planet, which decreases with stellar temperature as the habitable zone moves outwards towards the peak of the giant planet occurrence rate.}
    \label{fig:chz2_steff}
\end{figure}

The CHZ$_2$ metric from \citet[][]{Ware2025}, calculated for the 164 stars on the EMSL, tended to increase with increasing stellar temperature. To first order, the spatial extent of the HZ increases as $(\frac{L}{L_\odot})^{0.5}$. For the HZ models used here, the radiative transfer of higher effective temperature stellar spectra through planetary atmospheres also results in somewhat wider HZs at a given instellation, though this is a smaller effect. The increase in HZ width with stellar effective temperature dictates a corresponding increase in the peak CHZ$_2$ metric. However, this increase cuts off at $\gtrsim 6500$ Kelvin. F stars hotter than this tend to evolve too quickly and not spend enough time on the main sequence to maintain a stable habitable zone for 2 billion years (see Figure~\ref{fig:chz2_steff}). \citet[][]{Ware2025} found that three systems (HD 90089 A, HD 90589, and HD 109085) have a CHZ$_2$ metric of 0, indicating that life present on any habitable planets in these systems may not have had enough time to make a detectable impact on the atmosphere. All three systems had a small ($<1\%$) chance of hosting a temperate terrestrial planet via our simulations.

Additionally, there were 35 systems where the probability of temperate terrestrial planets clashed with the CHZ$_2$ metric. We define said clashes in one of two ways. The first is if the probability for finding a temperate terrestrial planet is relatively high ($>0.1$) but the CHZ$_2$ is relatively low ($< 0.1$). The second is if the CHZ$_2$ is relatively high ($> 0.5$) but the probability of a temperate terrestrial planet is relatively low ($<0.01$). In this manner, 30 systems had a very low probability of hosting rocky temperate planets but a high CHZ$_2$ metric, whereas 5 systems had a relatively higher probability of finding a rocky planet in the habitable zone but a low CHZ$_2$ metric. For the 30 systems with low probability but high CHZ$_2$, we note that 8 of these systems currently contain planets, most of which preclude the probability of an additional planet in the system being rocky and in the circumstellar habitable zone. Of the remaining 22, all but one were spectral type G2 or earlier, which shows that F stars are likely to have a strong CHZ$_2$, but more likely to also have giant planets in or near their habitable zones based on current occurrence rates, likely precluding the formation of rocky planets.

We also note that, in general, the EMSL similarly ranks early-G and F stars lower than later type stars. The tiers of the EMSL list were determined primarily by circumstellar disk presence, lack of close binarity, and apparent size of the habitable zone for the coronagraph limit. HZ planets around earlier type stars will be further from the host star due to the $(\frac{L}{L_\odot})^{0.5}$ scaling. This results in lower planet-star brightness ratios with increasing stellar effective temperature and higher coronagraph contrast ratio requirements. Also, there are simply more later-type K and late-G stars close to the Sun, where the apparent size of the habitable zone is larger because the star is closer, thus increasing the rankings of the nearby later-type stars. However, the EMSL neglects the effect of stellar effective temperature on HZ width, which is nearly six times wider for early F stars than late K stars \citep[1.7 AU vs 0.3 AU; e.g.,][]{Kopparapu2014}. More importantly, a geometric albedo of 0.2 was assumed for all EMSL test cases, but the albedo of a planet with an Earth-like or CO$_2$ dominated atmosphere will increase with increasing stellar effective temperature \citep[][]{kasting1993}. For one, the Rayleigh scattering cross-section is proportional to $\lambda^{-4}$, and so, given Wien's Law, $\sigma_R \propto T_{\rm eff}^4$. This leads to increased diffuse reflection as the stellar spectrum is blue-shifted. The absorption coefficients for H$_2$O and CO$_2$ are also weaker in the visible than in the near-infrared, leading to less incident light being absorbed for planets with hotter host stars. These unaccounted for factors will counter some of the effect of further HZ distances by increasing the proportion of reflected light. \cite{delgenio2019} used an ensemble of general circulation models to examine the change in Bond albedo with respect to stellar $T_{\rm eff}$ and instellation for a variety of habitable planets. They found that for more weakly irradiated planets ($0.94 \geq S/S_\oplus \geq 0.32$), where the Bond albedo is more sensitive to changes in $T_{\rm eff}$, the Bond albedo increases by $\sim 0.06$ for each 1000 K increase in $T_{\rm eff}$. We note that the EMSL adopted the geometric albedo of 0.2 from the exo-Earth yield modeling work of \cite{stark2014}, but the authors explicitly state that albedos may be one of the largest sources of uncertainty in yield modeling.

Finally, \citet[][]{Dencs2025} found that around 1-10\% of moons around Jovian planets at 1-2 AU from a G-type host star could potentially be habitable. Our simulations agree with these measures, as we find about 1-10\% of the Jovian planets we create fall in the 1-2 AU range, and overall we see a 0.01-1\% chance of finding a habitable Earth-like moon across all simulated or known systems.

\section{Conclusions}{\label{sec:conclusions}}

In this study we attempted to create a prioritization ranking system for the 164 stellar systems on the EMSL, which are currently the highest ranked tier of stars to observe with the Habitable Worlds Observatory. This ranking system is based upon the predicted probability of temperate terrestrial planets or moons of giant planets, where the potential for long-term habitability is determined via the known exoplanet demographics, dynamical stability considerations, and heat flux calculations. If we are able to optimize target selection, we will likely be able to increase the yield and observing time of temperate terrestrial exoplanets studied via direct imaging and spectroscopy to search for biosignatures in the atmospheres of these planets. The main conclusions from our analysis are as follows:

\begin{itemize}
    \item There are no systems on the EMSL currently known to host a rocky planet in the habitable zone, out of the $\sim20\%$ known to host any planets. Thus, any ranking of the stars in the list by priority will have to be done with simulated planets or systems.
    \item For the stars with no known planets, by creating a simulated system with Kepler ``peas-in-a-pod" chains of small planets interior to a giant planet, we find that the K stars on the list have the highest probability of finding a temperate terrestrial planet. This is likely due to small planets being more likely found in the habitable zone of a K star, given the population demographics of small planets and truncation due to outer giant planets.
    \item However, we find that the continuous habitable zone metric tends to devalue later-type stars and favor earlier G and F type stars due to the larger stars having correspondingly larger habitable zones, which have a lower probability of terrestrial planets due to the increase in the giant planet occurrence rate in their habitable zones.
    \item For stars with known planets, we find that $\sim40\%$ of them are extremely unlikely to host a temperate terrestrial planet, supporting a similar analysis by \citet[][]{Kane2024}, whereas $\sim20\%$ reach significant probabilities ($>15\%$).
    \item For systems with giant planets within or exterior to the circumstellar habitable zone, we created simulated moon systems to augment the possibility of finding a temperate terrestrial body, but on average less than $0.1\%$ of created moons were potentially habitable.
\end{itemize}

HWO is planned to be a multi-billion-dollar observatory that will transform our search for life outside of our solar system. Thus, prioritizing its ability to find planets around the stars most likely to host the desired targets is important in order to avoid any potential misses of temperate terrestrial planets. Additional work to observe and model exoplanet systems will further improve these rankings and provide HWO with a more robust set of targets to search for signs of life.

\begin{acknowledgements}
    The authors would like to thank Eric Mamajek, Caleb Harada, and Nuri Park for informative discussions that improved the content of the manuscript. The results reported herein benefited from collaborations and/or information exchange within NASA’s Nexus for Exoplanet System Science (NExSS) research coordination network sponsored by NASA’s Science Mission Directorate (grant 80NSSC23K1356, PI: Steve Desch).
\end{acknowledgements}

\begin{contribution}
    JD created the statistical models, performed a majority of the analyses, and wrote most of the manuscript. KR performed the remaining analyses, helped define one of the models, and edited the manuscript. AW assisted in stellar parameters in the models, wrote sections of the manuscript pertaining to the target stars and continuous habitable zones, and edited the manuscript. PY contributed to defining the hypotheses tested with the analyses and edited the manuscript.
\end{contribution}

\software{\texttt{Astropy} \citep[][]{Astropy2022}, \texttt{DYNAMITE} \citep[][]{Dietrich2020}, \texttt{SciPy} \citep[][]{Virtanen2020}, \texttt{TidalPy} \citep[][]{Renaud2023}}

\appendix

\section{Systems with no known planets} {\label{app:simulated}}

\begin{longtable}{c|c|c|c|c|c}
    \centering
    \textbf{System} & $\mathbf{\hat{\eta_\oplus}}$ & $\mathbf{CHZ_2}$ & \textbf{Moon} & \textbf{\textit{Joint Metric}} & \textbf{EMSL Tier}\\
    \hline
HD 201091 & \{0.2872, 0.386\} & 0.323 & \{0.00487, 0.00796\} & 0.1108 & A\\
HD 201092 & \{0.3889, 0.5258\} & 0.205 & \{0.00458, 0.00739\} & 0.095 & A\\
HD 165341 B & \{0.2783, 0.3751\} & 0.263 & \{0.005, 0.00824\} & 0.0877 & B\\
HD 131977 & \{0.1838, 0.2552\} & 0.343 & \{0.00451, 0.00783\} & 0.0774 & A\\
HD 88230 & \{0.3523, 0.4764\} & 0.176 & \{0.00487, 0.00801\} & 0.0741 & A\\
HD 202560 & \{0.379, 0.5121\} & 0.162 & \{0.00465, 0.00788\} & 0.0732 & A\\
HD 191408 A & \{0.1426, 0.2011\} & 0.388 & \{0.0052, 0.00839\} & 0.0693 & C\\
HD 155886 & \{0.1284, 0.1824\} & 0.400 & \{0.005, 0.00823\} & 0.0648 & B\\
HD 155885 & \{0.1276, 0.1817\} & 0.400 & \{0.00496, 0.00818\} & 0.0645 & B\\
HD 26965 A & \{0.0941, 0.1374\} & 0.451 & \{0.00502, 0.00823\} & 0.0552 & A\\
HD 156026 & \{0.2628, 0.3536\} & 0.174 & \{0.00487, 0.00831\} & 0.0548 & A\\
HD 4628 & \{0.1332, 0.1896\} & 0.313 & \{0.00484, 0.00813\} & 0.0526 & A\\
HD 10476 & \{0.0782, 0.1168\} & 0.469 & \{0.00515, 0.0085\} & 0.0489 & A\\
HD 185144 & \{0.0878, 0.1281\} & 0.424 & \{0.00496, 0.00824\} & 0.0486 & A\\
HD 165341 A & \{0.0673, 0.0997\} & 0.491 & \{0.00537, 0.00895\} & 0.0445 & B\\
HD 43834 A & \{0.0596, 0.0759\} & 0.578 & \{0.00561, 0.00927\} & 0.0435 & C\\
HD 131156 A & \{0.0703, 0.1034\} & 0.462 & \{0.00554, 0.00844\} & 0.0433 & B\\
HD 216803 & \{0.2173, 0.2945\} & 0.159 & \{0.00522, 0.00822\} & 0.0418 & B\\
HD 10360 & \{0.1087, 0.1567\} & 0.299 & \{0.00537, 0.00853\} & 0.0418 & A\\
HD 109358 & \{0.0501, 0.0641\} & 0.634 & \{0.0063, 0.00979\} & 0.0413 & A\\
HD 101501 & \{0.0636, 0.0927\} & 0.480 & \{0.0055, 0.00864\} & 0.0409 & A\\
HD 156274 A & \{0.0853, 0.1243\} & 0.365 & \{0.00493, 0.00833\} & 0.0407 & B\\
HD 38392 & \{0.1515, 0.2121\} & 0.215 & \{0.00458, 0.00763\} & 0.0404 & B\\
HD 10361 & \{0.1303, 0.1858\} & 0.245 & \{0.00511, 0.00859\} & 0.0404 & A\\
HD 32147 & \{0.1304, 0.1862\} & 0.244 & \{0.00537, 0.00869\} & 0.0403 & B\\
HD 82885 A & \{0.063, 0.0829\} & 0.499 & \{0.00571, 0.00927\} & 0.0401 & B\\
HD 1581 & \{0.0461, 0.0601\} & 0.638 & \{0.00589, 0.00939\} & 0.0387 & B\\
HD 100623 A & \{0.1091, 0.1592\} & 0.265 & \{0.00529, 0.00842\} & 0.0374 & A\\
HD 20807 & \{0.0574, 0.0729\} & 0.510 & \{0.00565, 0.00926\} & 0.037 & A\\
HD 128621 & \{0.0763, 0.1134\} & 0.361 & \{0.00524, 0.0087\} & 0.0368 & B\\
HD 20630 & \{0.0602, 0.0768\} & 0.479 & \{0.00595, 0.00916\} & 0.0364 & A\\
HD 190248 & \{0.0374, 0.0499\} & 0.672 & \{0.00627, 0.0099\} & 0.0347 & A\\
HD 10780 & \{0.0718, 0.1056\} & 0.363 & \{0.00487, 0.00804\} & 0.0346 & A\\
HD 131156 B & \{0.3124, 0.4219\} & 0.088 & \{0.00489, 0.00797\} & 0.0329 & C\\
HD 128620 & \{0.033, 0.0445\} & 0.690 & \{0.00666, 0.01109\} & 0.0329 & B\\
HD 203608 & \{0.0382, 0.0505\} & 0.621 & \{0.00613, 0.00995\} & 0.0325 & A\\
HD 149661 & \{0.0796, 0.1176\} & 0.307 & \{0.00509, 0.00821\} & 0.0323 & A\\
HD 20766 & \{0.0612, 0.0833\} & 0.398 & \{0.00565, 0.00925\} & 0.0317 & A\\
HD 146233 & \{0.0494, 0.0642\} & 0.438 & \{0.00603, 0.0096\} & 0.0283 & A\\
HD 157214 & \{0.036, 0.0476\} & 0.539 & \{0.00597, 0.00983\} & 0.0268 & C\\
HD 34411 & \{0.0245, 0.034\} & 0.710 & \{0.00622, 0.01037\} & 0.0267 & A\\
HD 114710 & \{0.0412, 0.0537\} & 0.456 & \{0.00604, 0.00979\} & 0.0252 & A\\
HD 17925 & \{0.0912, 0.1338\} & 0.210 & \{0.00524, 0.00898\} & 0.0251 & B\\
HD 50281 & \{0.2038, 0.2787\} & 0.097 & \{0.00481, 0.00804\} & 0.024 & C\\
HD 33262 A & \{0.0398, 0.0523\} & 0.437 & \{0.00605, 0.00961\} & 0.0235 & B\\
HD 84117 & \{0.0216, 0.03\} & 0.686 & \{0.00646, 0.01036\} & 0.0234 & A\\
HD 35296 & \{0.0284, 0.0387\} & 0.546 & \{0.00657, 0.01032\} & 0.023 & A\\
HD 207129 & \{0.0478, 0.0618\} & 0.353 & \{0.00582, 0.00927\} & 0.022 & C\\
HD 65907 A & \{0.0445, 0.058\} & 0.366 & \{0.00617, 0.01027\} & 0.0218 & B\\
HD 55575 & \{0.0378, 0.0501\} & 0.393 & \{0.00707, 0.01117\} & 0.0209 & B\\
HD 182572 & \{0.0185, 0.027\} & 0.672 & \{0.00608, 0.00994\} & 0.0207 & C\\
HD 48682 & \{0.025, 0.0346\} & 0.532 & \{0.00632, 0.01\} & 0.0202 & C\\
HD 165499 & \{0.0254, 0.0357\} & 0.516 & \{0.00643, 0.01006\} & 0.02 & B\\
HD 30495 & \{0.0569, 0.0732\} & 0.269 & \{0.00533, 0.00895\} & 0.0194 & B\\
HD 53705 & \{0.0324, 0.0431\} & 0.414 & \{0.0061, 0.01\} & 0.019 & B\\
HD 122064 & \{0.1296, 0.1866\} & 0.114 & \{0.00482, 0.00797\} & 0.0188 & C\\
HD 90839 & \{0.0343, 0.0464\} & 0.374 & \{0.00644, 0.01034\} & 0.0182 & A\\
HD 103095 & \{0.2329, 0.3131\} & 0.064 & \{0.00483, 0.00811\} & 0.0179 & C\\
HD 69897 & \{0.013, 0.0205\} & 0.680 & \{0.00661, 0.01067\} & 0.0173 & B\\
HD 19373 & \{0.0104, 0.0167\} & 0.760 & \{0.00675, 0.0109\} & 0.017 & A\\
HD 46588 A & \{0.0276, 0.0385\} & 0.394 & \{0.00654, 0.01048\} & 0.0164 & B\\
HD 4391 & \{0.0593, 0.0759\} & 0.213 & \{0.00519, 0.00863\} & 0.0159 & B\\
HD 140538 A & \{0.0633, 0.0801\} & 0.198 & \{0.00548, 0.0094\} & 0.0157 & B\\
HD 50692 & \{0.0439, 0.0571\} & 0.264 & \{0.0058, 0.00931\} & 0.0153 & B\\
HD 7570 & \{0.0197, 0.0287\} & 0.473 & \{0.00608, 0.01004\} & 0.0153 & B\\
HD 160915 & \{0.0105, 0.017\} & 0.663 & \{0.00679, 0.01076\} & 0.0149 & C\\
HD 16895 A & \{0.0167, 0.0246\} & 0.455 & \{0.00679, 0.01097\} & 0.0134 & A\\
HD 37394 & \{0.0756, 0.112\} & 0.131 & \{0.00532, 0.00895\} & 0.0132 & B\\
HD 166 & \{0.0629, 0.0895\} & 0.152 & \{0.0056, 0.00876\} & 0.0127 & B\\
HD 72905 & \{0.0591, 0.0748\} & 0.163 & \{0.00622, 0.00961\} & 0.0122 & B\\
HD 219623 & \{0.0201, 0.0285\} & 0.365 & \{0.00657, 0.01058\} & 0.012 & C\\
HD 187691 A & \{0.0026, 0.0076\} & 0.771 & \{0.00687, 0.01109\} & 0.0109 & B\\
HD 38393 & \{0.0148, 0.022\} & 0.392 & \{0.00696, 0.01137\} & 0.0108 & A\\
HD 38858 & \{0.0616, 0.0839\} & 0.134 & \{0.00585, 0.00949\} & 0.0108 & C\\
HD 58855 & \{0.0101, 0.0165\} & 0.499 & \{0.00629, 0.01019\} & 0.0108 & B\\
HD 193664 & \{0.052, 0.0666\} & 0.148 & \{0.00623, 0.01017\} & 0.01 & C\\
HD 125276 A & \{0.0491, 0.0632\} & 0.150 & \{0.00546, 0.00934\} & 0.0095 & C\\
HD 110897 & \{0.0541, 0.0691\} & 0.136 & \{0.00614, 0.00988\} & 0.0095 & C\\
HD 4614 A & \{0.0463, 0.0606\} & 0.154 & \{0.00573, 0.00939\} & 0.0094 & A\\
HD 4813 & \{0.0291, 0.0396\} & 0.206 & \{0.00684, 0.01069\} & 0.0089 & A\\
HD 114837 A & \{0.0007, 0.0052\} & 0.758 & \{0.00651, 0.01097\} & 0.0088 & C\\
HD 142860 & \{0.0, 0.0041\} & 0.792 & \{0.00676, 0.01121\} & 0.0088 & A\\
HD 32923 & \{0.0018, 0.007\} & 0.648 & \{0.00663, 0.01063\} & 0.0085 & C\\
HD 76151 & \{0.0566, 0.0717\} & 0.115 & \{0.0059, 0.00945\} & 0.0083 & C\\
HD 102870 & \{0.0, 0.0\} & 0.904 & \{0.00684, 0.01129\} & 0.0082 & C\\
HD 126660 A & \{0.0, 0.0\} & 0.883 & \{0.00694, 0.01151\} & 0.0081 & C\\
HD 5015 & \{0.0, 0.0\} & 0.859 & \{0.00682, 0.01128\} & 0.0078 & C\\
HD 78154 A & \{0.0, 0.0\} & 0.865 & \{0.0064, 0.01079\} & 0.0074 & C\\
HD 222368 & \{0.0, 0.0\} & 0.794 & \{0.00694, 0.01163\} & 0.0074 & C\\
HD 693 & \{0.0, 0.0016\} & 0.725 & \{0.00659, 0.01058\} & 0.0068 & A\\
HD 22484 & \{0.0, 0.0\} & 0.745 & \{0.00703, 0.01117\} & 0.0068 & B\\
HD 74576 & \{0.1176, 0.1696\} & 0.045 & \{0.00501, 0.00823\} & 0.0068 & C\\
HD 160032 & \{0.0, 0.0\} & 0.758 & \{0.00665, 0.01109\} & 0.0067 & C\\
HD 168151 & \{0.0, 0.0\} & 0.804 & \{0.00626, 0.0104\} & 0.0067 & C\\
HD 20010 A & \{0.0, 0.0\} & 0.716 & \{0.00686, 0.01152\} & 0.0066 & C\\
HD 84737 & \{0.0, 0.0\} & 0.740 & \{0.00655, 0.01112\} & 0.0065 & B\\
HD 23249 & \{0.0, 0.0\} & 0.707 & \{0.00685, 0.01081\} & 0.0062 & A\\
HD 215648 A & \{0.0, 0.0\} & 0.701 & \{0.00664, 0.01104\} & 0.0062 & C\\
HD 2151 & \{0.0, 0.0\} & 0.698 & \{0.00649, 0.01117\} & 0.0062 & C\\
HD 187013 & \{0.0, 0.0\} & 0.670 & \{0.00678, 0.0112\} & 0.006 & C\\
HD 91324 & \{0.0, 0.0\} & 0.684 & \{0.00599, 0.01034\} & 0.0056 & C\\
HD 142373 & \{0.0, 0.0\} & 0.597 & \{0.00688, 0.0112\} & 0.0054 & A\\
HD 206860 & \{0.0501, 0.0649\} & 0.082 & \{0.00589, 0.00949\} & 0.0053 & C\\
HD 72673 & \{0.096, 0.1401\} & 0.039 & \{0.00515, 0.00802\} & 0.0049 & C\\
HD 212330 A & \{0.0, 0.0\} & 0.550 & \{0.00639, 0.01045\} & 0.0046 & B\\
HD 14412 & \{0.0935, 0.137\} & 0.037 & \{0.00521, 0.00884\} & 0.0045 & C\\
HD 7788 A & \{0.0, 0.0\} & 0.455 & \{0.00669, 0.01091\} & 0.004 & C\\
HD 165185 & \{0.0541, 0.0702\} & 0.055 & \{0.00609, 0.00965\} & 0.0038 & C\\
HD 158633 & \{0.095, 0.1388\} & 0.031 & \{0.00529, 0.00844\} & 0.0038 & C\\
HD 17206 & \{0.0077, 0.0138\} & 0.157 & \{0.00652, 0.01074\} & 0.003 & A\\
HD 25457 & \{0.0231, 0.0324\} & 0.064 & \{0.00678, 0.0108\} & 0.0023 & C\\
HD 78366 & \{0.0448, 0.0584\} & 0.037 & \{0.00611, 0.00977\} & 0.0022 & C\\
HD 210302 & \{0.0024, 0.0074\} & 0.136 & \{0.00662, 0.01099\} & 0.0019 & A\\
HD 156897 A & \{0.0, 0.0\} & 0.199 & \{0.00677, 0.01124\} & 0.0018 & C\\
HD 89449 & \{0.0, 0.0\} & 0.186 & \{0.00682, 0.01114\} & 0.0017 & C\\
HD 199260 & \{0.0254, 0.0351\} & 0.041 & \{0.00638, 0.0103\} & 0.0016 & C\\
HD 128167 & \{0.0042, 0.0093\} & 0.089 & \{0.00744, 0.01191\} & 0.0015 & B\\
HD 105452 A & \{0.0, 0.0\} & 0.119 & \{0.00609, 0.01057\} & 0.001 & C\\
HD 219482 & \{0.029, 0.0388\} & 0.023 & \{0.0064, 0.0102\} & 0.001 & C\\
HD 739 & \{0.0039, 0.009\} & 0.059 & \{0.00699, 0.01135\} & 0.0009 & C\\
HD 25998 & \{0.0153, 0.0223\} & 0.026 & \{0.00652, 0.01089\} & 0.0007 & C\\
HD 134083 & \{0.0, 0.0032\} & 0.064 & \{0.00667, 0.01093\} & 0.0007 & B\\
HD 64379 & \{0.015, 0.0222\} & 0.013 & \{0.00662, 0.01056\} & 0.0004 & C\\
HD 43386 & \{0.0052, 0.0107\} & 0.018 & \{0.00675, 0.01079\} & 0.0003 & B\\
HD 22001 A & \{0.0, 0.0\} & 0.017 & \{0.00659, 0.011\} & 0.0001 & C\\
HD 43042 & \{0.0049, 0.0104\} & 0.005 & \{0.00674, 0.01096\} & 0.0001 & B\\
HD 197692 & \{0.0, 0.0\} & 0.007 & \{0.00667, 0.01105\} & 0.0001 & C\\
HD 30652 & \{0.0077, 0.0134\} & 0.003 & \{0.0068, 0.01089\} & 0.0001 & A\\
HD 213845 A & \{0.0, 0.0013\} & 0.001 & \{0.00667, 0.01052\} & 0.0 & C\\
HD 23754 & \{0.0, 0.0\} & 0.001 & \{0.0061, 0.00995\} & 0.0 & C\\
HD 90589 & \{0.0, 0.0\} & 0.000 & \{0.00637, 0.01064\} & 0.0 & C\\
HD 109085 & \{0.0, 0.0\} & 0.000 & \{0.00643, 0.0108\} & 0.0 & C\\
HD 90089 A & \{0.0031, 0.0078\} & 0.000 & \{0.00652, 0.01065\} & 0.0 & C\\
HD 166620 & \{0.0, 0.0\} & 0.070 & \{0.0, 0.0\} & 0.0 & C\\
    \caption{The three individual metrics and the joint metric for each of the 134 stars on the EMSL without known planets. $\hat{\eta_\oplus}$ is the probability of finding an Earth-like planet within the habitable zone in our simulations, as in Table~\ref{tab:metrics_known}. The conservative and optimistic cases are shown separately for the probability of the temperate terrestrial planet and moon, whereas the joint metric uses the average value in its calculation.}
    \label{tab:metrics_unknown}
\end{longtable}

%% For this sample we use BibTeX plus aasjournalv7.bst to generate the

\bibliography{main}{}
\bibliographystyle{aasjournalv7}



\end{document}